\documentclass[twocolumn,amsmath,amssymb,showpacs]{revtex4}
\usepackage{graphicx}

\begin{document}

\title{Superconductivity in doped inversion-symmetric Weyl semimetals}

\author{Tao Zhou$^{1,2}$}
\email{tzhou@nuaa.edu.cn}

\author{Yi Gao$^{3}$}

\author{Z. D. Wang$^{2}$}
\email{zwang@hku.hk}

\affiliation{$^{1}$College of Science, Nanjing University of Aeronautics and Astronautics, Nanjing 210016, China.\\
$^{2}$Department of Physics and Center of Theoretical and Computational Physics, The University of Hong Kong, Pokfulam Road, Hong Kong, China.\\
$^{3}$Department of Physics and Institute of Theoretical Physics,
Nanjing Normal University, Nanjing, 210023, China.}

\date{\today}
\begin{abstract}
We study theoretically the superconductivity in doped Weyl semimetals with an inversion symmetry based on the Bogoliubov-de
Gennes equations. In principle, the two  superconducting states, i.e., the zero momentum BCS-like pairing and the finite momentum Fulde-Ferrell-Larkin-Ovchinnikov (FFLO) pairing are competing in this kind of systems.
Our self-consistent calculation indicates that the BCS-type state may be the ground state. The competition between these two pairing states is studied in detail through normal state Fermi surface and the finite energy spectral functions. Generally, the Fermi surface topology supports the FFLO pairing while the finite energy band structure favors the BCS-type pairing.
We also study the physical properties and address the Majorana Fermions excitation in these two superconducting state respectively.

\end{abstract}
\maketitle

\section{introduction}

In the past decade, there has been significant interest in topological phases of condensed matter~\cite{qi}.
In fully gapped systems, the nontrivial topological phase is defined by a quantized topological invariant, which implies the
appearance of the edge states on the boundary of
the system. Recently, research  has also been extended to the topological gapless systems~\cite{wan,zhao}. In particular, the Weyl semimetal is such a representative topological fermionic gapless one~\cite{wan}, which
has pairs of gapless points (Weyl nodes) in the bulk spectrum and Fermi arcs on the boundaries, even for disordered ones~\cite{zhao2}.
An idea of Weyl semimetal in condensed matter system was first proposed theoretically that it may be realized in a class of pyrochlore iridates~\cite{wan}. Recently it was predicted that it may be realized in non-centrosymmetric transition metal
monophosphides (including  TaAs, TaP, NbAs and NbP) based on the first principle calculation~\cite{weng,huang}.
Soon after the theoretical prediction, the indications of Weyl fermions in TaAs~\cite{syxu,bqlv}, NbAs~\cite{sxu}, TaP~\cite{nxu,yxu}, and NbP~\cite{wang}, have been reported experimentally. Another kind of Weyl semimetal is the WX$_2$-family material (WTe$_2$ and WoTe$_2$)~\cite{sol,ysun}. Very interestingly, it was revealed that WoTe$_2$ material is a superconductor with the maximum transition temperature at 8.2 K under pressure~\cite{yqi}, providing us with a potential platform to study the interplay between the topology and superconductivity in Weyl semimetal systems.

Generally the superconducting state should be the ground state of a metal via the pairing interaction mediated by the phonon or other bosons. Thus it is understandable to observe the superconductivity in Weyl semimetals. Actually,  the superconductivity in Weyl semimetal systems has been studied
before~\cite{meng,cho,das,chen,wei,kha,yang,bed,liu,blu,jian}. Theoretically the ground state of the Weyl-type superconductors is still to be confirmed yet. Starting from the lattice model, it was proposed that
a finite momentum pairing state [Fulde-Ferrell-
Larkin-Ovchinnikov (FFLO) state] is favorred over the uniform zero momentum pairing (BCS-type pairing)~\cite{cho} .
While an odd-parity superconductivity is proposed by later studies~\cite{wei,bed}. It was proposed that the FFLO state is favorred for local interaction while the BCS state wins out for non-local interaction~\cite{wei}. Very recently, it was indicated that the BCS state is the ground state for an inversion symmetric doped Weyl semimetals~\cite{bed}.
 Note that all of the above studies are merely based on a free energy analysis. The gap magnitude is self-consistently determined while the form of the gap functions is preset. To our knowledge,  full self-consistent studies about the superconducting pairing in this kind of systems are still awaited.
Also, for both BCS and FFLO states, there exist some exotic features, such as the crossed flat
bands in the uniform BCS-type pairing state~\cite{blu}, and the spacetime supersymmetry in the FFLO-type pairing state~\cite{jian}. Thus it is timely and of fundamental interest to study the competition of these two superconducting states in more detail and clarify what kind of state is the genuine ground state.

Another important issue in topological superconducting systems is the excitations of the Majorana Fermions (MFs), which can in principle help us to realize non-abelian statistics and have potential applications in topological quantum computation~\cite{naya}. Generally in a bulk gapped topological superconductor, the MFs naturally appear at the system boundaries. While for the Weyl superconductors, the BCS-type pairing state is bulk gapless, different from usual topological superconductors. And so far, the numerical identification for MFs in topological FFLO states is also lack. Thus it is insightful to identify the MFs numerically in possible superconducting states in Weyl semimetals.

In this paper, motivated by the above considerations, we study numerically the ground state and the MFs in the Weyl superconductors from the lattice-type model with the inversion symmetry. Following Refs.~\cite{cho,bed}, we here consider the superconductivity in doped Weyl semimetals.
While different from their studies which are based on the free energy analysis, we here perform a self-consistent study based on the
the Bogoliubov-de
Gennes (BdG) equations. Our numerical results reveal that the BCS pairing state is favorable. The competition between the BCS state and the FFLO state is studied in detail through the normal state Fermi surface and the spectral functions. The physical properties in these two superconducting states are also investigated. The MFs in both the BCS state and the FFLO state are studied numerically.

The rest of the paper is organized as follows.
In Sec. II, we introduce
the model and work out the formalism. In Sec. III, we
perform numerical calculations and discuss the obtained
results. Finally, we give a brief summary in Sec. IV.

\section{Model and formalism}

We start from a model Hamiltonian including the normal state term and the pairing term,
\begin{equation}
H=H_t+H_{SC}.
\end{equation}
The normal state term $H_t$ is a general two-band model, given by
\begin{eqnarray}
H_t=&-\sum\limits_{ {\bf i}\alpha\sigma}\sigma t_\alpha (c^{\dagger}_{{\bf i}\sigma} c_{{\bf i+\hat{\alpha}}\sigma}+h.c.)+\sum_{{\bf i}\sigma}(\sigma h-\mu)c^{\dagger}_{{\bf i}\sigma} c_{{\bf i}\sigma}\nonumber\\
&+\sum_{\bf i}(i\lambda c^{\dagger}_{{\bf i}\uparrow}c_{{\bf i}+\hat{x}\downarrow}+i\lambda c^{\dagger}_{{\bf i}\downarrow}c_{{\bf i}+\hat{x}\uparrow}+h.c.\nonumber\\
&+\lambda c^{\dagger}_{{\bf i}\uparrow}c_{{\bf i}+\hat{y}\downarrow}-\lambda c^{\dagger}_{{\bf i}\downarrow}c_{{\bf i}+\hat{y}\uparrow}+h.c.),
\end{eqnarray}
with ${\bf i}=(x,y,z)$ represents a site on the three dimensional cubic lattice. $\hat{\alpha}=\hat{x}$, $\hat{y}$, or $\hat{z}$ represents
the base vector along $x$, $y$, or $z$ direction.

$H_{SC}$ is the on-site superconducting pairing term, expressed by
\begin{equation}
H_{SC}=\sum_{\bf i}(\Delta_{\bf ii}c^{\dagger}_{{\bf i}\uparrow}c^{\dagger}_{{\bf i}\downarrow}+h.c.).
\end{equation}

The above Hamiltonian can be diagonalized by solving the Bogoliubov-de
Gennes (BdG) equations,
\begin{equation}
\sum_{\bf j}\left(
\begin{array}{cccc}
 H_{{\bf ij}\uparrow\uparrow} & H_{{\bf ij}\uparrow\downarrow} & \Delta_{{\bf jj}} & 0 \\
 H_{{\bf ij}\downarrow\uparrow} & H_{{\bf ij}\downarrow\downarrow} & 0 & -\Delta_{{\bf jj}}\\
 \Delta^{*}_{{\bf jj}} & 0 & -H_{{\bf ij}\downarrow\downarrow} & -H^{*}_{{\bf ij}\downarrow\uparrow}\\
0 & -\Delta^{*}_{{\bf jj}} & -H^{*}_{{\bf ij}\uparrow\downarrow} & -H_{{\bf ij}\uparrow\uparrow}
\end{array}
\right)  \begin{array}{c} \Psi^{\eta}_{\bf j}
\end{array}
 =E_\eta \begin{array}{c}\Psi^{\eta}_{\bf j}
\end{array},
\end{equation}
where
$\Psi^{\eta}_{\bf j}= (u^{\eta}_{{\bf j}\uparrow},u^{\eta}_{{\bf j}\downarrow},v^{\eta}_{{\bf
j}\downarrow},v^{\eta}_{{\bf j}\uparrow})^{\mathrm{T}}$.
$H_{{\bf ij}\sigma\sigma}$ and $H_{{\bf ij}\sigma\bar{\sigma}}$ $(\sigma\neq \bar{\sigma})$ are obtained from Eq.(2).
The superconducting order parameter $\Delta_{\bf jj}$ is calculated
self-consistently,
\begin{eqnarray}
\Delta_{\bf jj}=\frac{V}{2}\sum_\eta u^{\eta}_{{\bf
j}\uparrow}v^{\eta*}_{{\bf j}\downarrow}\tanh (\frac{E_\eta}{2K_B T}),
\end{eqnarray}
with $V$ being the pairing strength.

The edge states of the system may be studied with a
cylindrical geometry, i.e., considering the periodic boundary condition along the $x$ direction.
Thus the Hamiltonian can be reduced to a quasi-two-dimensional one by a partial Fourier transformation,
\begin{equation}
c_{\underline{\bf i}\sigma}(k_x)=\frac{1}{\sqrt{L_x}}\sum_{x}c_{\underline{\bf i}\sigma}e^{ik_x x},
\end{equation}
where $L_x$ is the period of the lattice along the $x$ direction, and $\underline{\bf i}=(y,z)$ represents a site in the reduced $yz$-plane.
As a result, the normal state Hamiltonian $H_t$ may be rewritten as
\begin{eqnarray}
H_t=&-\sum\limits_{k_x\underline{\bf i}\alpha\sigma}\sigma t_\alpha [c^{\dagger}_{\underline{\bf i}\sigma}(k_x) c_{{\underline{\bf i}+\hat{\alpha}}\sigma}(k_x)+h.c.]\nonumber\\
&+\sum\limits_{k_x\underline{\bf i}\sigma}(\sigma h-2\sigma t_x\cos k_x-\mu)c^{\dagger}_{\underline{\bf i}\sigma} (k_x)c_{\underline{\bf i}\sigma}(k_x)\nonumber\\
&+\sum\limits_{k_x\underline{\bf i}}[\lambda c^{\dagger}_{\underline{\bf i}\uparrow}(k_x)c_{\underline{\bf i}+\hat{y}\downarrow}(k_x)-\lambda c^{\dagger}_{\underline{\bf i}\downarrow}(k_x)c_{\underline{\bf i}+\hat{y}\uparrow}(k_x)\nonumber\\&+2\lambda \sin k_x c^{\dagger}_{\underline{\bf i}\uparrow}(k_x)c_{\underline{\bf i}\downarrow}(k_x)+h.c.].
\end{eqnarray}

In the superconducting state, in the mean-field level,  both uniform BCS type and FFLO type solutions are in principle possible when the spin-polarized term exists. Usually the FFLO modulation is suppressed and the BCS state is the ground state with a strong spin-orbital interaction~\cite{zhou}.
In the present model, a strong spin-flip hopping term is considered in the $x$ direction, which could suppress the FFLO modulation along this direction. Thus
it is reasonable to consider the superconducting order parameter being uniform along the $x$ direction.
This is qualitatively consistent with the normal state Fermi surface analysis in doped Weyl semimetals.
Generally the Fermi surface should consist of two disconnected Fermi pockets around the pair of Weyl points~\cite{cho}. In principle, two competing pairing states are possible, i.e., the inter-pocket BCS pairing and the intra-pocket FFLO pairing. Note that for both pairing states the net momentum of the Cooper pair along the $k_x$ direction keeps zero [see Fig.~1 in Ref.~\cite{cho}].
That is, the superconducting pairing term can be written as,
\begin{equation}
H_{SC}=\sum_{\underline{\bf i}}[\Delta_{\underline{\bf i}\underline{\bf i}}c^{\dagger}_{\underline{\bf i}\uparrow}(k_x)c^{\dagger}_{\underline{\bf i}\downarrow}(-k_x)+h.c.].
\end{equation}

The $k_x$ dependent BdG equation can be obtained from Eqs.(7) and (8), with the formalism similar to Eq.(4). The order parameter $\Delta_{\underline{\bf j}\underline{\bf j}}$ is calculated as,
\begin{equation}
\Delta_{\underline{\bf jj}}=\frac{V}{2L_x}\sum_{k_x\eta} u^{\eta}_{\underline{\bf
j}\uparrow}(k_x)v^{\eta*}_{\underline{\bf j}\downarrow}(k_x)\tanh (\frac{E_\eta(k_x)}{2K_B T}).
\end{equation}

In the reduced low-dimensional system, we may define a spectral function depending on the site and partial momentum $A_{\underline{\bf i}\sigma}({\bf k},\sigma)$ as
\begin{equation}
A_{\underline{\bf i}}({\bf k},\omega)=\sum_{\eta,\sigma} \frac{u^{\eta}_{\underline{\bf i}\sigma}({\bf k})^2}{\omega-E_\eta({\bf k})+i\Gamma}.
\end{equation}

The normal state Hamiltonian [Eq.(2)] can be expressed in the momentum space by a full Fourier transformation, which is written as the $2\times2$ matrix,
\begin{equation}
H_t=(h-2\sum_\alpha t_\alpha \cos k_\alpha)\sigma_z +2\lambda (\sin k_x \sigma_x + \sin k_y \sigma_y)-\mu \sigma_0.
\end{equation}
Here $\sigma_0$ is the identity matrix and $\sigma_{x,y,z}$ are the Pauli matrices. The two normal state energy bands are given by
\begin{equation}
E({\bf k})=\mu\pm \sqrt{4\lambda^2(\sin^2k_x+\sin^2k_y)+(h-2\sum_\alpha t_\alpha \cos k_\alpha)^2}.
\end{equation}
With $\mu=0$, the system may enter a Weyl semimetal phase.
A pair of Weyl points $W_{\pm}$ may be obtained from the above energy bands through choosing appropriate parameters, with
$W_{\pm}=(0,0,\pm \arccos [\frac {h-2 t_x-2t_y}{2 t_z}])$.

In the present work, the parameters are chosen as: $t_x=t_y=0.5$, $t_z=1$, $\lambda=0.5$, $h=2+2\cos(\pi/4)$. In this case, there exist two Weyl points at $(0,0,\pm\pi/4)$.  With the chemical potential $\mu=0.5$, the system may be metalic with a Fermi pocket surrounding each Weyl point. We have checked numerically the our main results are not sensitive to the parameters.

\section{Results and Discussions}

We first study the partially Fourier transformed Hamiltonian from Eqs.(7) and (8).
The BdG equations are solved self-consistently and the superconducting order parameters are obtained based on Eq.(9).
The superconductivity appears with a strong attractive interaction $(V\geq 7)$, qualitatively consistent with previous calculations~\cite{jian}.
However, here only uniform superconducting order parameters are obtained, i.e., the BCS-type pairing is supported by a self-consistent calculation. We also checked numerically that this result is qualitatively the same when the input parameters and initial order parameters are varied.
In detail, we consider different initial input order parameters: uniform ones (BCS-type), periodic ones (FFLO-type), and random ones. The converged solution is always uniform when the self-consistency is achieved.

The starting model of the present work is qualitatively the same as that in Ref.~\cite{cho}, while it was indicated there that an FFLO-type superconducting should be a most favorable state.
It is interesting to note that  their result is opposite to ours. Another difference is that in Ref.~\cite{cho} superconductivity can occur for very weak interaction, which is also inconsistent with our numerical calculation.

 \begin{figure}
\centering
  \includegraphics[width=3in]{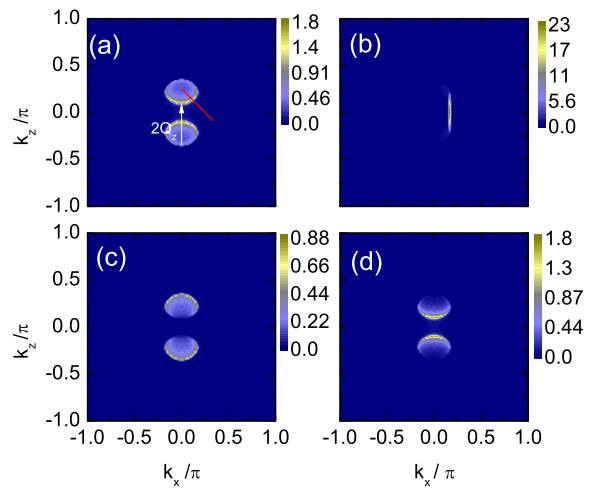}
\caption{(Color online) Intensity plots of normal state zero energy spectral functions with open boundary condition along $y$ direction ($1\leq y\leq 200$). (a) The spectral function in the bulk with y=100. The arrow indicates the Fermi surface nesting vector. (b) The spectral function at the system edge with y=1. (c) The spectral function of spin up particles with y=100. (d)  The spectral function of spin down particles with y=100.
}
\end{figure}

We now give an explanation for these disagreements. Actually, the appearance of the FFLO state is understood based on the Fermi surface topology.
We consider the Hamiltonian in $k_x-y-k_z$ space with open boundary condition along $y$-direction ($1\leq y\leq200$). The zero energy spectral functions $[A_y(k_x,k_z,\omega=0)]$
in the bulk and the system edge are presented in Figs.~1(a) and 1(b), respectively.
 As is known, $A_y(k_x,k_z,0)$ should be maximum at the Fermi momentum. Thus the normal state Fermi surface is obtained from Fig.~1.
In the bulk, as is seen in Fig.~1(a), there exists one Fermi pocket surrounding each Weyl point $(0,0,\pm Q_z)$, with the size of the pockets controlled by the chemical potential $\mu$.
At the system edge, as is shown in Fig.~1(b), open Fermi arc forms, which is connecting the tips of the bulk Fermi pockets and is
perpendicular to the
$k_x$-axis. The spin dependent zero energy spectral functions $A({\bf k},\uparrow)$ and $A({\bf k},\downarrow)$ are plotted in Figs.~1(c) and 1(d), respectively.
As is shown, significant spin imbalance exists: the spin up quasiparticles have large spectral weight as $\mid k_z\mid>\pi/4$, while the spin down ones have large spectral weight as $\mid k_z\mid<\pi/4$.
And it seems that the Fermi surface nesting exists with the nesting vector $2Q_z=\pi/2$, as indicated in Fig.~1(a).
As a result, if the superconductivity is completely determined by the quasiparticles on the Fermi surface,
 the pairing between two quasiparticles with the momentum ${\bf k}$ and ${-\bf k}\pm {\bf Q_f}$ should be favorable, with ${\bf Q_f}=(0,2Q_z)$.
In Ref.~\cite{cho}, the superconducting pairing is considered merely at a thin shell around the Fermi surface.
 Thus the superconducting pairing is determined entirely by the Fermi surface topology. As a result, the FFLO-type pairing is supported and the superconductivity occurs for even very weak attractive interactions.

However, for the present model, it is needed to emphasize that the Fermi surface analysis is not enough to confirm the superconducting pairing because the Fermi pockets are too tiny~\cite{note}.
Now let us study what kind of pairing is more favorable in more detail. We first illustrate
the two possible pairing states in Fig.~2(a). For BCS pairing state, the quasiparticles momenta from one pair are ${\bf k}$ and ${\bf -k}$, respectively. And for the FFLO state, the momenta are ${\bf k}$ and $-{\bf k}+{\bf Q_f}$, respectively. So more conclusive results may be obtained by comparing the spin dependent energy bands at the momenta ${\bf k}$, $-{\bf k}$, and $-{\bf k}+{\bf Q_f}$.
For the present model, the energy bands from the BdG equations are usually the superpositions of the spin up and spin down electrons due to the spin flip term. It is more insightful to study the spectral functions to obtain the spin dependent quasiparticle energies.

\begin{figure}
\centering
  \includegraphics[width=3in]{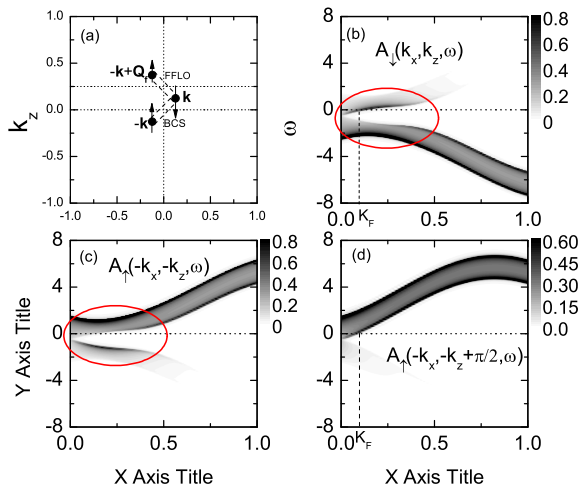}
\caption{(Color online)(a) Schematical illustration of the BCS pairing and FFLO pairing, respectively. (b-d) The spin-dependent spectral functions as functions of the energy and momentum with $k_z=\pi/4-k_x$ [along the (red) solid line cut shown in Fig.~1(a)].
}
\end{figure}

The spectral function for the spin down quasiparticles as functions of the energy and momentum [$A_{\downarrow}({\bf k},\omega)$] is plotted in Fig.~2(b).
The corresponding spectral functions for spin up quasiparticles at the momenta $-{\bf k}$ and $-{\bf k}+{\bf Q_f}$ are presented in Figs.~2(c) and 2(d), respectively.
We first compare Figs.~2(b) with 2(c) to roughly evaluate whether the BCS pairing is favorred.
At low energies, as displayed,  the energy bands at ${\bf k}$ and ${-\bf k}$ are obviously similar. The quasiparticle velocities are small near the Fermi energy from both spectra. The increase of the density of states for low energy quasiparticles may increase the possibility of the BCS pairing. We then compare Figs.~2(b) with 2(d) to study the possibility of the FFLO pairing. As is seen, the Fermi momenta $K_F$ are nearly the same. This feature favors the FFLO pairing. While, on the other hand, their quasiparticle band dispersions are significantly different, that is, the states at $-{\bf k}+{\bf Q_f}$ and ${\bf k}$ have different energies when away from the Fermi surface. This feature is unfavourable to the FFLO pairing.

We can estimate the volume of the two Fermi pockets from Fig.~1 or Eq.(12),
which is only about $0.35 \%$ of that of the Brillouin zone.
Thus for the present case,
the Fermi pockets are very tiny and the finite energy quasiparticles are also important to determine the ground states. As a result, when considering the quasiparticle pairing in the whole Brillouin zone, the BCS pairing wins over the FFLO pairing.

Very recently, it was also proposed that BCS-type pairing should win out for an inversion-symmetric doped Weyl semimetal system~\cite{bed}.
This is consistent with our above analysis.
While in Ref.~\cite{bed}, the odd-parity pairing state is revealed, consistent with a
previous study~\cite{wei}. Note that the lattice-type model is considered in
the present work, qualitatively different from those in Refs.~\cite{wei,bed}.
For a lattice model,  the odd parity pairing would obviously not occur for the on-site pairing potential.
The superconductivity in the odd-parity channel may be studied by taking into account a nearest-neighbour pairing potential. This is an interesting issue and may require further investigation.

 We now study the physical properties in the BCS-type superconducting state with the uniform pairing order parameter $\Delta_{\bf ii}\equiv \Delta_0$. After a full fourier transformation, it is found that the system has gapless nodes. In detail, if the condition $\mid h-2 t_x- 2t_y\pm \sqrt{\mu^2+\Delta_0^2}\mid< 2t_z$ is satisfied, there are four nodes at points $(0,0,\pm Q_\pm)$, with
 \begin{equation}
 Q_\pm=\arccos \frac{h-2 t_x- 2t_y\pm \sqrt{\mu^2+\Delta_0^2}}{2 t_z}.
 \end{equation}
As the chemical potential $\mu$ or pairing order $\Delta_0$ increases, the nodes $(0,0,\pm Q_{+})$ may disappear and only the nodes $(0,0,\pm Q_{-})$ are left.
As a result, there may exist two different BCS states, named as BCS-I state with four bulk nodes and BCS-II state with two bulk nodes.
Further increasing the pairing strength, the nodes $\pm Q_{-}$ may also disappear while this may require unreasonably large superconducting order parameter. This case is not considered in the present work.

\begin{figure}
\centering
  \includegraphics[width=3in]{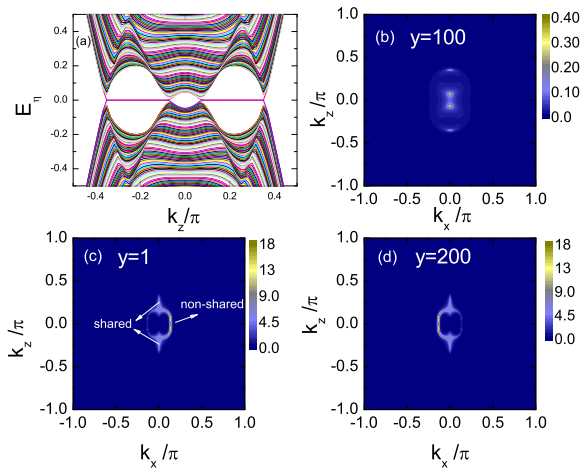}
\caption{(Color online) Numerical results of the BCS-I state. (a) The eigenvalues of the Hamiltonian along $k_x=0$. (b-d) Intensity plots of
zero energy spectral functions.
}
\end{figure}

We first consider the physical properties in BCS-I state with $\Delta_0=0.2$. This gives $Q_{+}$ and $Q_{-}$ to be $0.07 \pi$ and $0.36\pi$, respectively.
By considering the open boundary condition along the $y$-direction and
transforming the Hamiltonian to the momentum space along $x$ and $z$ directions,
the quasiparticle energy spectra along $k_x=0$ is presented in Fig.~3(a).
There exists two segments of Fermi arcs connecting nodal points $Q_{+}$ and $Q_{-}$, indicating the presence of the edge states. The zero energy spectral functions in the bulk and system edges are plotted in Figs.~3(b-d). As is seen, in the bulk four Fermi points at the momentum $k_z=\pm Q_{\pm}$ exist, consistent with our above analysis and the band structure shown in Fig.~3(a).
At the system edges, there exist several segments of Fermi arcs. Especially, the Fermi arcs connecting the Fermi points $(0,Q_{-})$ and $(0,Q_{+})$ [or $(0,-Q_{-})$ and $(0,-Q_{+})$] have the same spectral weight at the two system boundary, as presented in Fig.~3(c) and 3(d). Such shared edge states is similar to that of the Majorana bound states and one zero energy quasiparticle may be decoupled by two spatially separated one. At the momentum $-Q_{+}< k_z <Q_{+}$, there exist non-shared edge states, with $k_x<0$ and $k_x>0$ parts belong to different boundaries.

It is important to pinpoint that there is no separate MFs in the BCS-I state, even when the shared edge states exist.
Usually two MFs $\gamma_{1,2}$ can be obtained from one zero mode $C=u_{{k}}\psi^\dagger({k})+v_{-k}\psi(-{k})$ with $\gamma_1=C+C^\dagger$ and $\gamma_2=i(C-C^\dagger)$.
For usual topological superconductors, generally the edge states occurs at the high-symmetric points, i.e., $k=0$ or $k=\pi$.
As a result, $k$ and $-k$ are equivalent points. Then $\gamma_1$ and $\gamma_2$ are naturally decoupled and locating at different boundaries.  While in the present BCS-I case , it is obvious that one can not obtain the separate $\gamma_{1,2}$ through the above procedure.


\begin{figure}
\centering
  \includegraphics[width=3in]{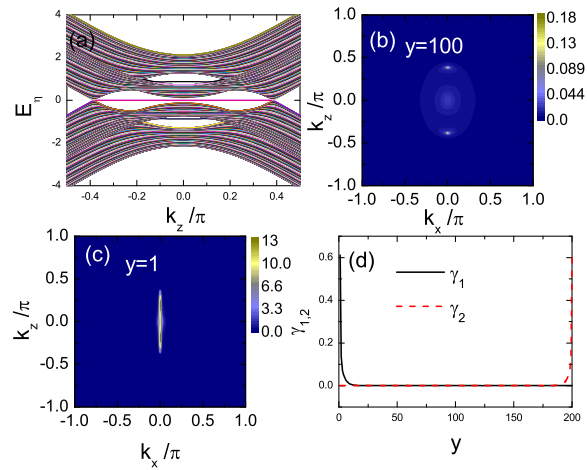}
\caption{(Color online) Numerical results of the BCS-II state. (a) The eigenvalues of the Hamiltonian along $k_x=0$. (b) and (c) are intensity plots of
zero energy spectral functions at the system bulk and edge, respectively. (d) The spatial distributions of the two MFs at the momentum $(k_x,k_z)=(0,0)$.
}
\end{figure}

We now turn to look into the properties of the BCS-II state. Generally this state can be achieved by increasing the chemical potential $\mu$ or the pairing magnitude $\Delta_0$. We here consider the pairing magnitude $\Delta_0=0.5$, which generates the two bulk nodes $(0,0,\pm Q_{-})$ with $Q_{-}=0.38\pi$.
The numerically results for this state are presented in Fig.~4. The energy spectrum along $k_x=0$ is plotted in Fig.~4(a). The zero energy spectral functions in the bulk and at system edge are presented in Figs.~4(b) and 4(c), respectively. From Fig.~4(a), there are two nodal points in the bulk spectrum and one segment of Fermi arc connecting $Q_{-}$ and $-Q_{-}$ at the system edge.
Here the Fermi arc is shared by the two system boundaries.
A significant different result from the BCS-I state is that there is a zero energy edge state at the momentum $(0,0)$ point, gives the possibility of the MFs excitation. The existence of the MFs excitation is further confirmed numerically. The spatial distribution of the two MFs from the zero energy fermion at $(0,0)$ point is displayed in Fig.~4(d). As is seen, two MFs are completely separated and locate at the two system edges.

\begin{figure}
\centering
  \includegraphics[width=3in]{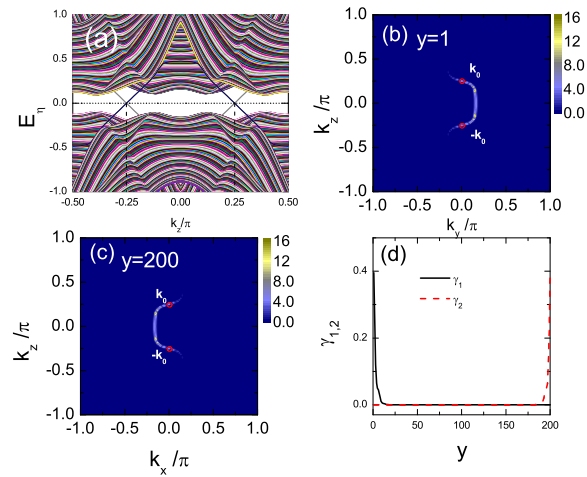}
\caption{(Color online) Numerical results of the FFLO-type pairing. (a) The eigenvalues of the Hamiltonian along $k_x=0$. (b-c) Intensity plots of
zero energy spectral functions at two boundaries, respectively. (d) The spatial distributions of the two MFs at the momentum ${\bf k_0}$.
}
\end{figure}

At last we would like to look into the FFLO-type pairing state. Although the FFLO-type pairing is not supported by our self-consistent calculation.
While it may be favorred by the Fermi surface analysis, thus it is still a potential pairing state and may appear upon some additional interaction. The FFLO-type
pairing is denoted as
\begin{equation}
H_{FFLO}=\sum_{\bf i}(\Delta_1 e^{i{\bf Q_f} \cdot  {\bf R_i}}c^{\dagger}_{{\bf i}\uparrow}c^{\dagger}_{{\bf i}\downarrow}+\Delta_2 e^{-i{\bf Q_f} \cdot  {\bf R_i}}c^{\dagger}_{{\bf i}\uparrow}c^{\dagger}_{{\bf i}\downarrow}+h.c.).
\end{equation}
The above Hamiltonian can be rewritten in the momentum space,
\begin{equation}
H_{FFLO}=\sum_{\bf k}(\Delta_1 c^{\dagger}_{{\bf k}\uparrow}c^{\dagger}_{{\bf -k}+{\bf Q_f}\downarrow}+\Delta_2 c^{\dagger}_{{\bf k}\uparrow}c^{\dagger}_{{\bf -k}-{\bf Q_f}\downarrow}+h.c.).
\end{equation}
Taking into account the inversion symmetry, we here consider the LO state with $\Delta_1=\Delta_2=0.2$.

The numerical results for the FFLO state are presented in Fig.~5.
The quasiparticle energies along $k_x=0$ is plotted in Fig.~5(a). There exist two zero energy states at the momentum $k_z=\pm \pi/4$. The
bulk energy spectrum is fully gapped and no Fermi surface exists.
 The zero energy spectral functions at the system edges are displayed in Figs.~5(b) and 5(c), respectively. The open Fermi arcs exist at the system edges. The Fermi arcs are
 not shared by the two boundaries, except for the points $\pm{\bf k_0}$ with ${\bf k_0}=(0,\pi/4)$, which are the crossing points of the two Fermi arcs.
At the points $\pm{\bf k_0}$, the quasiparticles have the same spectral weight at the two boundaries.
  Thus the MFs excitation may occur at these two points. Here ${\bf k_0}$ and the net Cooper pair momentum ${\bf Q_f}$ satisfies: ${\bf k_0}\equiv{\bf Q_f}/2$. This is an important relation. It is robust and does not change upon parameters.

The shared edge state at the momentum ${\bf k_0}$ is essential to produce MFs.
 In the FFLO state, the quasiparticles can be expressed as $C=u_{{{\bf k}}}\psi^\dagger({\bf k})+v_{{-\bf k}+{\bf Q_f}}\psi(-{\bf k}+ {\bf Q_f})$. At the point ${\bf k}={\bf k_0}$, the quasiparticle is expressed as $C=u_{{{\bf k_0}}}\psi^\dagger({\bf k_0})+v_{\bf k_0}\psi({\bf k_0})$. Obviously the quasiparticle is a MF if the condition $u_{{{\bf k_0}}}=v^{*}_{\bf k_0}$ satisfies. Actually, one can obtain two MFs $\gamma_{1,2}$ according to the standard method. Our numerical results verify that two separate MFs indeed exist. The numerical result for the spatial distribution of $\gamma_{1,2}$ is plotted in Fig.~5(d). As is seen, two MFs locate at two boundaries of the system.

\section{summary}

 In summary, we have studied theoretically the superconductivity in inversion symmetric doped Weyl semimetals based on the BdG equations. Our self-consistent calculations have indicated that the BCS-like pairing may be more favorable than the FFLO pairing. The competition between the BCS-like pairing and FFLO pairing is discussed based on the Fermi surface topology and finite energy band structure. The FFLO pairing is supported by the former and the BCS-like pairing is favorred by the latter.
The physical properties of the BCS-like pairing and FFLO pairing states have also been addressed.
Two different BCS states, named BCS-I state and BCS-II state, are revealed. There are four bulk nodes in BCS-I state and no MFs exists in this state. For the BCS-II state, the number of bulk nodes reduces to two and the MFs excitation occur in this state. For the FFLO state, the energy spectrum is full gapped. The open Fermi arc appears at the system edge. The separate MFs exist at the system boundaries.

\
\begin{acknowledgments}
 We thank D. B. Zhang and Y. X. Zhao for helpful discussions. This work was supported
by the NSFC (Grant No. 11374005), the NCET (Grant No. NCET-12-0626), Jiangsu Qingnan engineering project, the GRF (Grant Nos.  HKU173051/14P and HKU173055/15P), and the CRF (Grant No. HKU8/11G) of Hong Kong.
\end{acknowledgments}



\begin{thebibliography}{References}
\bibitem{note} For tiny Fermi pockets, the topologically-nontrivial
properties are qualitatively the same with those in undoped Weyl semimetals, while
the toplogical features will be broken for large Fermi pockets.
\end{thebibliography}
\end{document}